\documentstyle[11pt,fleqn]{article}
\newcommand{\dip}{\smallskip\it Dipartimento di Fisica, Universit\`a di Trento,
                      Italia}
\newcommand{\infn}{\smallskip\it Istituto Nazionale di Fisica Nucleare,
                       Gruppo Collegato di Trento, Italia}
\newcommand{\dinfn}{\dip \mbox{ } and \\ \infn}
\newcommand{\email}[1]{Email: \sl #1@itnvax.cineca.it,
                              #1@itncisca.bitnet, itnvax::#1}
\newcommand{\utf}[1]{\hfill{\sl U.T.F. #1}\par\medskip\par}
\newcommand{\abs}[1]{\hrule\par\begin{description}\item{Abstract: }
                     \it #1\par\end{description}\hrule\par\medskip}
\newcommand{\pacs}[1]{\noindent{\sl PACS numbers:
                       \hspace{0.3cm}#1}\par\bigskip}

\newcommand{\ack}[1]{\par\section*{Acknowledgments} #1}
\newcommand{\ca}[1]{{\cal #1}}         
\newcommand{\hs}{\hspace{2cm}}         
\newcommand{\nn}{\nonumber}            
\newcommand{\ap}{\left.}               
\newcommand{\at}{\left(}               
\newcommand{\aq}{\left[}               
\newcommand{\cp}{\right.}              
\newcommand{\ct}{\right)}              
\newcommand{\cq}{\right]}              
\newcommand{\beq}{\begin{equation}}                    
\newcommand{\eeq}{\end{equation}}                      
\newcommand{\beqn}{\begin{eqnarray}}                   
\newcommand{\eeqn}{\end{eqnarray}}                     
\newcommand{\ii}{\infty}                         
\newcommand{\X}{\times}                          
\newcommand{\fr}[2]{\mbox{$\frac{#1}{#2}$}}      
\newcommand{\tr}{\,\mbox{tr}\,}                  
\newcommand{\Tr}{\,\mbox{Tr}\,}                  
\newcommand{\ach}{\,\mbox{cosh$^{-1}$}\,}        
\newcommand{\lap}{\triangle}                     
\newcommand{\al}{\alpha}

\newcommand{\ga}{\gamma}
\newcommand{\de}{\delta}

\newcommand{\ze}{\zeta}

\newcommand{\ka}{\kappa}
\newcommand{\la}{\lambda}
\newcommand{\ro}{\varrho}

\newcommand{\Ga}{\Gamma}

\newcommand{\Om}{\Omega}

\begin{document}

\title{The trace of the heat kernel \\
on a compact hyperbolic 3-orbifold}

\author{Guido Cognola\thanks{\email{cognola}}
\mbox{} and Luciano Vanzo\thanks{\email{vanzo}}
\\  \dinfn}
\date{March 1993}
\maketitle
\utf{288}

\abs{The heat coefficients related to the Laplace-Beltrami
operator defined on the hyperbolic compact manifold
$H^3/\Ga$ are evaluated in the case in which
the discrete group $\Ga$ contains elliptic and hyperbolic elements.
It is shown that while hyperbolic elements give only exponentially
vanishing corrections to the trace of the heat kernel,
elliptic elements modify all coefficients of the asymptotic expansion,
but the Weyl term, which remains unchanged.
Some physical consequences are briefly discussed in the examples.}


\bigskip

In the last decades, there has been a great deal of investigations
on the properties of interacting quantum field theories in curved
space-time.
Several techniques have been employed. Among these, we mention
the background-field method within path-integral
approach \cite{dewi65b}, which is very
useful in dealing with the one-loop approximation and which permits
to evaluate the one-loop effective action.
As a consequence, all physical interesting quantities
(one-loop effective potential, vacuum energy, quantum anomalies and
so on) can be derived in a straightforward way.

The one-loop effective action, as derived from the path-integral,
is an ill defined quantity, being related to the determinant
of the fluctuation operator. Many regularization schemes have been
proposed. One of the most promising, which works very well in a
curved manifold too, is the so called ``zeta-function regularization''
\cite{hawk77-55-133},
by which one can define the determinant
of the fluctuation operator and therefore
the regularized one-loop effective action.

On a generic Riemannian manifold $\ca{M}$,
the zeta-function related to the
Laplace-Beltrami operator is generally unknown,
but nevertheless, some physical interesting quantities
can be related to its Mellin inverse transform,
which has a computable asymptotic expansion (heat kernel expansion).
The situation is really better on manifolds with constant curvature
where the effective action is explicitly known whenever the topology is
trivial, i.~e.~$\pi_{1}(\ca{M})=\{e\}$.
Where it not, the answer is generally unknown, but in the case of
compact hyperbolic manifolds,
the use of Selberg trace formula and Selberg
zeta-function give many informations about the effective action.

Quite recently, there have been attemps to investigate the role
of topology in quantum field theories and the phenomenon of
topological mass generation has been discovered
\cite{ford79-70-89,ford80-21-933,toms80-21-2805,toms80-129-334}.
The non trivial topological space-times which mainly have been
considered are the ones with the torus or spherical topology
(for a recent review see \cite{camp90-196-1} and references cited
therein).
It has to be noted however that compact hyperbolic manifolds have a
far reacher topological structure with respect to the previous ones
and for this reason they may provide many solutions with
interesting features for the construction of anisotropic
models of our universe. This is the main motivation for studying
quantum fields on space-times with constant curvature
spatial section and non trivial topology \cite{elli71-2-7}.

Now, the standard way to induce a non-trivial topology in a connected
Riemannian manifold $\ca{M}$, is to choose a discrete group of
isometries $\Ga$ of $\ca{M}$
acting effectively and properly discontinuously on $\ca{M}$ with
compact quotient space $\overline\ca{M}=\ca{M}/\Ga$.
This is still a topological
space but not necessarily a Riemannian manifold \cite{scot83-15-401}.
This is because the
group may have fixed points, in which case the induced metric will have
singularities at these points. For example, the quotient of
$R^{2}$ by the ciclic
group generated by a rotation of order $n$ is a cone with angle
$2\pi/n$, and the metric
has a singularity at the vertex of the cone.
If the group is generated by reflexions in a line $l$, then the quotient
is a half-plane whose boundary line is $l$ and again the metric is
singular at the fixed points ( lying on the line $l$). One can consider
also the group generated by a rotation of order $n$ about $0$ and reflexion
in a line through $0$, in which case the quotient is isometric to an
infinite wedge with angle $\pi/n$ with two semi-infinite boundary lines
of singular points.
In two dimensions, these are the only possible singularities for spaces
of this kind. Such spaces are called orbifolds,
after the work of Thurston \cite{thur82-6-357} which first
defined and studied them.
Explicit examples of orbifolds in two dimensions with physical
implications are given in refs.~\cite{dowk77-10-115,chan92r}.
In higher dimensions the situation
is not so simple and we refer to such singularities generically as
conical points, for brevity.
They naturally appear in finite
temperature field theories on the Euclidean section of space-times
which have an event horizon.
For example on de Sitter and Rindler space-times, conical
singularities are absent only if $T$ is equal to the
Hawking temperature \cite{furs93r}.

The trace of the heat-kernel of the Laplacian
on a compact hyperbolic 3-orbifold is
the main object of interest in this paper.
The relevance of the so called
Minakshisundaram-Schwinger-DeWitt expansion \cite{mina49-1-242,dewi65b}
of this kernel in the study of interacting quantum field
theories in curved space-time is well known.
Its role in renormalization theory is essential
(see for example the text book \cite{birr82b}
and referecences cited therein).

In the case of a second order-non negative-elliptic-symmmetric
differential operator $L$, defined on a $N$-dimensional
compact-Riemannian-smooth manifold ${\cal M}$ without boundary,
the asymptotic expansions for the operator $\Tr\exp(-tL)$ and for
its kernel $K_t(x,x;L)$, which is the fundamental solution of the heat
type equation

\begin{equation}
\partial_t\,K_t(x,x';L)+L_x\,K_t(x,x';L)=0;\hs
\lim_{t\to 0}K_t(x,x';L)=\de(x,x')
\end{equation}
read

\beqn
\Tr e^{-tL}&\sim&\sum_{n=0}^\ii A_n(L)\;t^{n-N/2}\nn\\
K_t(x,x;L)&\sim&\frac{\sqrt{g(x)}}{(4\pi)^{N/2}}
\sum _{n=0}^\ii a_n(x;L)\;t^{n-N/2}\\
A_n(L)&=&\frac{1}{(4\pi)^{N/2}}
\int_{{\cal M}}a_n(x;L)\sqrt{g}\;d^Nx\nn
\label{HKE}
\eeqn
where $g$ is the determinant of the metric tensor $g_{ij}$
with signature $N$ ($i,j=1,\dots,N$) and $\de(x)$ is the Dirac
delta function. By definition, $a_0(x;L)=1$.
The quantities $a_n(x;L)$ are the so called spectral
coefficients associated with $L$. They are invariant quantities related
to geometrical objects.
In general, their evaluation is a complicated task and only some of them
have been given in explicit form, using different techniques, as for example
Minakshisundaram-Schwinger-DeWitt anzatz \cite{mina49-1-242,dewi65b}
or pseudo-differential operator calculus
\cite{seel67-10-172,gilk75-10-601}.
Some years ago, they have been also computed in
the case of a Riemann-Cartan manifold with arbitrary
torsion \cite{obuk82-108-308,cogn88-3-599}.
We refer to the literature for general expressions and details.

When the manifold has a smooth boundary, the behaviour of the
expansion drastically changes. In the simplest case in which the
boundary $\partial{\cal M}$ is a $N-1$ dimensional smooth manifold,
it can be shown that, due to the singularity of $K_t(x,x;L)$ near the
boundary, $\Tr\exp(-tL)$ has an asymptotic expansion in which powers of
$\sqrt{t}$ appear too \cite{mcke67-1-43,grei71-41-163}, that is

\begin{equation}
\Tr e^{-tL}\sim\sum_{n=0}^\ii K_n(L)\;t^{(n-N)/2}
\end{equation}
Of course, in this case $K_n(L)$ are also related to the extrinsic
curvature of the boundary. It has to be noted, that the asymptotic for
the kernel $K_t(x,x;L)$ is the same which one as for the boundaryless case,
provided that $x\not\in\partial{\cal M}$. This is due to the fact
that all spectral coefficients depend on local properties of the manifold.
The computation of the spectral coefficients in the case with boundary is a
much more difficult task than the boundaryless case. In the literature both
classical methods
\cite{kac66-73-1,mcke67-1-43,bali70-60-401,bali71-64-261}
\cite{stew71-69-353,waec72-72-349,kenn78-11-173} as well as
pseudo-differential operator techniques
\cite{grei71-41-163,smit81-63-467,grub86b}
have been used. Quite recently, new methods have been proposed in
order to compute the spectral coefficients for manifolds with
boundary in refs.~\cite{bran90-15-245,cogn90-241-381}.
In particular, in ref.~\cite{bran90-15-245} the whole expression
for $K_2(L)$ can be found.

Exact solutions for the heat kernel of the operator $L=-\lap+X$,
$\lap$ being the Laplace-Beltrami operator and $X$ a constant,
exist on $R^N$, $S^N$, $H^N$ and on any homogeneous compact space
(for a recent review see ref.~\cite{camp90-196-1}
and references cited therin).

In the case of orbifolds, the existence of the asymptotic expansion
for the trace of the heat-kernel is a remarkable result of
H.~Donnelly \cite{donn76-224-161,donn79-23-485}.
Despite the metric singularities,
it is shown that the Laplace-Beltrami operator
has a self-adjoint extension on the orbifold with pure point spectrum
unbounded from above, but unlike than for manifolds, the smallest
eigenvalue is not necessarily zero.
The asymptotic expansion turns out to have the same form as for
manifolds but the coefficients are not given as polynomials in the
curvature tensor, which after all is undefined at the singularities.
The first coefficient was obtained in ref.~\cite{wall76-11-91}
and a general method
for computing the higher coefficients is also at our disposal
\cite{donn76-224-161}.
One important result is
$A_{0}=(4\pi)^{-N/2}\mbox{Vol}(\overline\ca{M})$,
so that the orbifold singularities do not affect the Weyl term.

For a compact hyperbolic 3-orbifold however, the Selberg trace formula
is sufficient to cover all cases of interest, so we turn our attention
to this case from now on. We shall obtain all coefficients
of the trace of the heat-kernel and consequently
we shall be able to discuss how the given
singularities affect some relevant physical quantities.

\bigskip

The hyperbolic space $H^3$ can be realized as the upper half-space in $R^3$,
that is $H^3\equiv\{P=(z,r)|z=x^1+ix^2\in C,r=x^3\in(0,\ii)\}$,
with the hyperbolic metric $ds^2=(d\bar zdz+dr^2)/r^2$.
The corresponding hyperbolic distance $\rho(P,P')$ is defined by

\begin{equation}
\cosh\rho(P,P')=\frac{|z-z'|^2+r^2+r'^2}{2rr'}
\end{equation}

The group  $SL(2,C)$ acts on $H^3$ in the following way. For
every matrix $\sigma\equiv([a,b],[c,d])\in SL(2,C)$ and every $P\in H^3$
one defines

\begin{equation}
\sigma P=\left(\frac{(az+b)(\bar c\bar z+\bar d)
+a\bar cr^2}{|cz+d|^2+|c|^2r^2},\frac{r}{|cz+d|^2+|c|^2r^2}\right)
\end{equation}

We shall consider the group of $H^3$ isometries
$PSL(2,C)=SL(2,C)/\{-1,1\}$, where 1 is the unity of $SL(2,C)$.
It is known that all elements of $PSL(2,C)$ belong to one of the
following conjugacy classes: elliptic, parabolic and hyperbolic
(loxodromic and strictly hyperbolic). The next step is to pick up
a discrete subgroup $\Ga \in PSL(2,C)$, acting properly
discontinuously on $H^{3}$ with compact quotient space $H^{3}/\Ga$.
Then $\Ga$ is called co-compact and it does not contain parabolic
elements
\cite{elst83-38-119,elst85p,elst87-277-655}.

The space $H^{3}/\Ga$ will contain lines along which the hyperbolic
metric is singular, due to the fact that the elliptic elements
of $\Ga$ fixes a line in $H^{3}$, namely the axis of a discrete rotation
( in two dimension they fix a point, which became a true conical
singularity in the quotient space).

Every hyperbolic element $T\in \Ga$ is conjugate to
a unique element $D(T)\in PSL(2,C)$ given by

\begin{equation}
D(T)=\left(
\begin{array}{cc}
a(T)&0\\0&a(T)^{-1}
\end{array}
\right) \\ |a(T)|>1
\end{equation}
The number $N(T)=|a(T)|^2$ is called the norm of $T$.
Likewise, every elliptic element $R\in\Ga$ is conjugate to a
unique element $D(R)\in PSL(2,C)$ given by

\begin{equation}
D(R)=\left(
\begin{array}{cc}
\xi(R)&0\\0&\xi(R)^{-1}
\end{array}
\right) \\ |\xi(R)|=1
\end{equation}

Let $C(T)$ be a centralizer of $T$ in $\Ga$, that is the set of
$\ga\in \Ga$ for which $\ga T=T\ga$ and $T_o\in C(T)$ an element
with minimal norm $N(T_o)>1$ among all elements of $C(T)$. Then
$T_o$ is called a primitive hyperbolic element for $T$
in $\Ga$. It is not uniquely determined by $T$ although $N(T_o)$ is
(for more details see
refs.~\cite{elst83-38-119,elst85p,elst87-277-655,gonc90-19-73}).
Similarly, the centralizer $C(R)$ of any elliptic element must contain
hyperbolic elements since $C(R)$ is proved to be infinite. Again we
denote by $T_{0}$ any element with minimal norm among the hyperbolic
elements of $C(R)$.
Let $\chi$ be an arbitrary finite-dimensional representation of $\Ga$
(a character of $\Ga$), i.~e.~a homomorphism $\chi:\Ga\to S^1$.
If $k(u)\in C_0^\ii([0,\ii))$ then the linear integral operator defined
by the kernel

\begin{equation}
K_{\Ga,\chi}(P,P')=\sum_{\ga\in\Ga}\chi(\ga)k(u(P,\ga P'))
\\P,P'\in H^3
\end{equation}
is of trace class in the standard Hilbert space and the equality

\begin{equation}
\sum_{j}h(r_j)=\int_{{\cal F}}K_{\Ga,\chi}(P,P)d\mu
=\sum_{\ga\in\Ga}\int_{{\cal F}}\chi(\ga)k(u(P,\ga P))d\mu
\label{hrj}
\end{equation}
holds
\cite{elst83-38-119,elst85p,elst87-277-655,venk82-4-1}.
Here $r_j$ is the positive square root of $\la_j+\ka$,
$\ka$ being the negative constant curvature of the manifold,
$\la_j$ are the eigenvalues of $-\lap$
(which contains finitely many eigenvalues less than $|\ka|$)
${\cal F}$ is a fundamental domain for $\Ga$,
$u(P,P')=(\cosh\rho(P,P')-1)/2$
is the two points invariant connected with the distance $\rho$
and $d\mu=r^{-3}dzd\bar zdr$ is the invariant measure.
As usual, for convenience we set $\ka=-1$. In this way all quantities
will be dimensionless.
When necessary, we will establish standard
units by a simple dimensional argument.
The pre-trace formula (9), is
actually valid in any dimension for strictly hyperbolic groups.

Then the function $k(u)$ is directly related to the Fourier
transform $\tilde h(p)$ of $h(r)$ by the formulae \cite{selb56-20-47}

\begin{equation}
\left\{
\begin{array}{l}
k(u)=(-4\pi)^{-n}\tilde h^{(n)}(p(u))\\
\tilde h(p(u))=(4\pi)^n\int_u^\ii du_1\int_{u_1}^\ii du_2
\cdots\int_{u_{n-1}}^\ii k(u_n)du_n
\end{array}
\right.\\ N=2n+1
\end{equation}

\begin{equation}
\left\{
\begin{array}{l}
2\pi k(u)=(-4\pi)^{-(n-1)}\int_u^\ii
\frac{\tilde h^{(n)}(p(z))dz}{\sqrt{z-u}}\\
2\pi \tilde h(p(u))=(4\pi)^{n}\int_u^\ii du_1\int_{u_1}^\ii du_2
\cdots\int_{u_{n-1}}^\ii \frac{k(u_n)du_n}{\sqrt{u_n-u_{n-1}}}
\end{array}
\right.\\ N=2n
\end{equation}
Here $p(z)=\ach(1+2z)$.
For $N=3$ then we have

\begin{equation}
k(u)=-\frac{1}{4\pi}\frac{d}{du}\tilde h(p(u))
=-\frac{1}{4\pi}\frac{d\tilde h(p)}{dp}\frac{dp(u)}{du}
\label{ku}
\end{equation}

The sum over $\ga$ in eq.~(\ref{hrj}) can be
expanded if one distinguishes between identity, elliptic and
hyperbolic elements. One has in fact the Selberg trace
formula \cite{elst85p}

\beqn
\sum_{j}h(r_j)&=& V({\cal F})k(0)+4\pi E\int_{0}^{\ii}k(u)du
+\sum_{\{T\}_{\Ga}} 4\pi H(T)\int_{\al(T)}^{\ii}k(u)du\nn\\
&=&V({\cal F})k(0)+E\tilde h(0)
+\sum_{\{T\}_{\Ga}} H(T)\tilde h(\log(N(T)))
\label{STF}
\eeqn
where

\begin{equation}
E=\sum_{\{R\}_{\Ga}}
\frac{\chi(R)\log N(T_o)}{m(R)|(\tr R)^2-4|}
\\
H(T)=\frac{\chi(T)\log N(T_o)}{m(T)|a(T)-a^{-1}(T)|^2}
\end{equation}
$E$ is a positive number called the elliptic number
of the manifold $H^3/\Ga$.
The summations are extended over the elliptic $\{R\}_{\Ga}$ and
hyperbolic $\{T\}_{\Ga}$ conjugacy classes in $\Ga$.
The numbers $m(R)$ and $m(T)$ are the orders of that minimal rotation
around the axis of $R$ and $T$ respectively, i.~e.~around the hyperbolic
line through the fixed point of $R$ and $T$ in $C\bigcup \{\ii \}$ and
are uniquely determined by $R$ and $T$, $T_{0}$ is any element with
minimal norm among the hyperbolic elements in $C(R)$ and
$C(T)$ respectively.
Finally, $\al(T)=[\cosh\log N(T)-1]/2$. A further partition of the
conjugacy classes into primitive ones is possible. We avoided this just
to keep formulas as simple as possible.

Now we use Selberg trace formula (\ref{STF}) to compute
$\Tr\exp(-tL_3)$ on $H^3/\Ga$, where $L_3=-\lap+X$, $X$ being a constant.
To this aim we choose the function $h(r)=\exp(-t(r^2+1))$. Its
Fourier transform reads

\begin{equation}
\tilde h(p)=\frac{1}{2\pi}\int_{-\ii}^{\ii}e^{-ipr}h(r)
=\frac{e^{(-t-p^2/4t)}}{\sqrt{4\pi t}}
\label{}
\end{equation}
from which

\begin{equation}
k(u)=\frac{p(u)p'(u)}{2}\frac{e^{(-t-p^2(u)/4t)}}{(4\pi t)^{3/2}} \\
k(0)=\frac{e^{-t}}{(4\pi t)^{3/2}}
\label{}
\end{equation}
directly follows.
Then eq.~(\ref{STF}) gives

\beqn
\Tr e^{-tL_3}&=&e^{-tX}\Tr e^{t\lap}\nn\\
&=&\aq V({\cal F})+4\pi tE\cq\frac{e^{-tM^2}}{(4\pi t)^{3/2}}
+\sum_{\{T\}_{\Ga}}
\frac{H(T)e^{-(tM^2+p^2(\al(T))/4t)}}{(4\pi t)^{1/2}}
\label{TrL3}
\eeqn
This is the exact expression valid for every group $\Ga$ with elliptic
and hyperbolic elements. We have set $M^2=X+1$ since in all formulae,
such a quantity plays the role of an effective mass. Of course we
suppose it to be strictly positive.
Looking at eq.~(\ref{TrL3}), one easily sees that the contribution
due to hyperbolic elements is exponentially vanishing when $t\to 0$.
This means that the last term in eq.~(\ref{TrL3}) does not contribute
to the parametrix of $\Tr\exp(-tL_3)$. Then, for small $t$ we have

\beq
\Tr e^{-tL_3}\sim \sum_{n=0}^{\ii}
\frac{e^{-tM^2}}{(4\pi t)^{3/2}} \aq V({\cal F})+4\pi t E\cq
\eeq
from which

\beq
A_n(L_3)=\frac{(-M^2)^n}{(4\pi)^{3/2}n!}
\aq V({\cal F})-\frac{4\pi nE}{M^2}\cq
\eeq
easily follows. As announced in the introduction, we see that
elliptic elements of the group $\Ga$, which realize the
compactification of $H^3$, modify all heat coefficients, but the first
one. This is due to the fact that the fixed points of elliptic
elements are singularities for the metric of $H^3/\Ga$.
Then the manifold stops to be smooth and
Gilkey theorem \cite{gilk75-10-601} is no longer valid.

The zeta-function related to $L_3$ can be easily computed by means of a
Mellin transform of expression (\ref{TrL3}). In fact one has

\begin{eqnarray}
\ze(s;L_3)&=&\frac{1}{\Ga(s)}\int_0^\ii t^{s-1}\Tr e^{-tL_3}\;dt\nn\\
&=&\frac{\Ga(s-3/2)}{(4\pi)^{3/2}\Ga(s)}
\aq V({\cal F})+\frac{4\pi(s-3/2)E}{M^2}\cq M^{-2(s-3/2)}\\
&&+\frac{\sin\pi s}{\pi}
\aq\int_{1}^{\ii}(u^2-1)^{-s}\frac{Z'}{Z}(1+uM)\;du\cq
M^{-2(s-1/2)}\nn
\label{zsL3}
\end{eqnarray}
where the logarithmic derivative of
Selberg $Z$-function on $H^3/\Ga$ is given by means of
formula \cite{gang77-21-1,elst85p}

\beq
\frac{Z'}{Z}(s)=\sum_{\{T\}_{\Ga}}
H(T)\;N(T)^{-(s-1)}
\eeq

\bigskip

As physical applications of the formulae we have just obtained,
we briefly consider a massive scalar field at finite temperature
$T$ defined on the
ultrastatic manifold ${\cal M}=R\X H^3/\Ga$ and evaluate
the vacuum energy and critical temperature of
Bose-Einstein condensation for that system.
Now we use standard units and so $\ka$ has the dimension of a square
mass.

The field operator is given by

\beq
L=-\lap+m^2+\xi R=-\frac{\partial^2}{\partial\tau^2}+L_3
\eeq
$m$ being the mass of the field and $\xi$ an arbitrary
dimensionless parameter.
It is interesting to observe that the scalar curvature $R$ of ${\cal M}$
is related to the curvature $\ka$ of $H^3$ by the equation
$R=6\ka$, and so

\beq
M^2=m^2+\xi R-\ka=m^2+(\xi-1/6)R
\eeq

The regularized vacuum energy is given by the
equation \cite{cogn92-33-222}

\beq
E_v=\frac{1}{2}\aq\bar\ze(-1/2;L_3)
-\frac{\bar\ell A_2(L_3)}{\sqrt{4\pi}}\cq
\eeq
where $\bar\ze(s_o;L_3)$ is the finite part of $\ze(s;L_3)$ in the
pole $s_o$ and $\bar\ell=2\log(e\ell/2)$, $\ell$ being an arbitrary
normalization parameter coming from the scalar path integral measure.
Using eq.~(\ref{zsL3}) we obtain \cite{cogn92-7-3677}

\begin{eqnarray}
E_v&=&\frac{V({\cal F})M^4}{64\pi^2}\at\log\fr{M^2}{\ell^2}-\fr{3}{2}\ct
-\frac{EM^2|\ka|^{-1/2}}{8\pi}\at\log\fr{M^2}{\ell^2}-\fr{1}{2}\ct
\nn\\
&&-\frac{M^2|\ka|^{-1/2}}{2\pi}
\int_{1}^{\ii}\sqrt{(u^2-1)}\frac{Z'}{Z}(1+uM|\ka|^{-1/2})\;du
\label{EvREG}
\end{eqnarray}
In order to fix the value of the renormalization parameter $\ell$,
we impose the vacuum energy density to be vanish in the
limit of zero curvature $\ka$. Recalling that
$V({\cal F})\propto |\ka|^{-3/2}$ one obtains
$\log(m^2/\ell^2)=3/2$ and so the renormalized vacuum energy reads

\begin{eqnarray}
E_{ren}&=&\frac{V({\cal F})M^4}{64\pi^2}\log\fr{M^2}{m^2}
-\frac{EM^2|\ka|^{-1/2}}{8\pi}\at\log\fr{M^2}{m^2}-1\ct
\nn\\
&&-\frac{M^2|\ka|^{-1/2}}{2\pi}
\int_{1}^{\ii}\sqrt{(u^2-1)}\frac{Z'}{Z}(1+uM|\ka|^{-1/2})\;du
\label{Evren}
\end{eqnarray}
It is interesting to observe that in the case of a conformally
invariant coupling (this means $m=0$ and $\xi=1/6$), $M^2=0$ and so
the vacuum energy is always vanishing.

\bigskip

Now we derive the critical temperature at which Bose-Einstein
condensation takes place. Strictly speaking, there is a sharp
critical temperature $T_c$ only in the case of infinite volume.
As usual \cite{park91-44-2421},
here we suppose the volume to be very large and compute the
``critical'' temperature taking the square of the
chemical potential $\mu$ equal to the smallest eigenvalue
of the operator $L$, which in our case is equal to $m^2+\xi R$.

For the thermodynamic potential $\Om(T,\mu)$
we have the high temperature expansion
\cite{cogn92-7-3677}

\begin{eqnarray}
\Om(T,\mu)&=&-\frac{V({\cal F})\pi^2}{45}T^4
+\left[\frac{V({\cal F})}{12}(M^2-2\mu^2)
-\frac{\pi E|\ka|^{-1/2}}{3}\right]T^2
\nn \\
&&-\aq \frac{V({\cal F})(M^2-\mu^2)^{3/2}}{6\pi}
-E|\ka|^{-1/2}(M^2-\mu^2)^{1/2}\cp\\
&&\ap +M|\ka|^{-1/2}\int_{1}^{\ii}\frac{u^2-1}{(\mu/M)^2+u^2-1}
\frac{Z'}{Z}(1+uM|\ka|^{-1/2})\;du
\cq T+\dots\nn
\label{HTE}
\end{eqnarray}
Taking the derivative of the latter expression
with respect to $-\mu$, we obtain an expansion for the charge density.
It reads

\begin{eqnarray}
\ro=\frac{\mu T^2}{3}
&-&\mu T\aq \frac{(M^2-\mu^2)^{1/2}}{2\pi}
-\frac{2E|\ka|^{-1/2}(M^2-\mu^2)^{-1/2}}{V({\cal F})}\cp
\\ && \phantom{\mu T}\ap
-\frac{2|\ka|^{-1/2}}{MV({\cal F})}
\int_{1}^{\ii}\frac{u^2-1}{[(\mu/M)^2+u^2-1]^2}
\frac{Z'}{Z}(1+uM|\ka|^{-1/2})\;du
\cq +\dots\nn
\end{eqnarray}

The critical temperature is given for $\mu^2=m^2+6\xi\ka$.
Recalling that the volume is proportional to $|\ka|^{-3/2}$ and taking
only leading terms into account, we finally obtain

\begin{equation}
T_c=\at\frac{3\ro}{m}\ct^{1/2}+\frac{3|\ka|^{1/2}}{4\pi}
-\frac{3E}{|\ka|V({\cal F})}+o(\ka)
\label{Tc}
\end{equation}
It has to be noticed, that in the limit of infinite volume
this is in agreement with the result of ref.~\cite{habe81-46-1497},
where Bose-Einstein condensation for a relativistic boson system was
considered in Minkowski space-time.
On the other and, eq.~(\ref{Tc})
is a little bit different from the analog result obtained for
$H^3$, since this is not the infinite volume limit
of $H^3/\Ga$. Rather, we are dealing with the limit of small curvature.

\ack{We would like to thank K.~Kirsten for helpful discussions
and suggestions}

\end{document}